 \documentclass[aps,prb,reprint,showpacs,amsmath,amssymb]{revtex4-1}
\usepackage{natbib}
\usepackage{float}
\usepackage[dvips,final]{graphicx}
\usepackage{color}
\usepackage{amsmath}
\usepackage{mathrsfs} % fancy hamiltonian

\usepackage{graphicx, subfigure}
\usepackage{pslatex}
\usepackage{relsize}
\usepackage{bm}
\usepackage{xspace} % nice spacing for math/text symbols

\usepackage{xcolor}
\usepackage{soul}

\def\ba{\begin{eqnarray}}
\def\ea{\end{eqnarray}}
\def\be{\begin{equation}}
\def\ee{\end{equation}}

\bibpunct{[}{]}{,}{n}{}{} % reset to online citations

\begin{document}
\title{New magnetic states in nanorings created by anisotropy gradients}

\author{M. Castro}
\author{A.P. Espejo}
%\author{R. Escobard}
\affiliation{Departamento de F\'isica, Universidad de Santiago de Chile and CEDENNA,\\ Avda. Ecuador 3493, Santiago, Chile}
\author{N.M. Vargas}
\affiliation{Department of Physics, University of California, San Diego, La Jolla, California 92093, United States}
\author{D. Altbir}
\author{S. Allende}
\affiliation{Departamento de F\'isica, Universidad de Santiago de Chile and CEDENNA,\\ Avda. Ecuador 3493, Santiago, Chile}
\author{V. L. Carvalho-Santos}
\email{vagson.carvalho@usach.cl}
\affiliation{Universidade Federal de Vi\c cosa, Departamento de F\'isica,
Avenida Peter Henry Rolfs s/n, 36570-000, Vi\c cosa, MG, Brasil}
\date{\today}

\begin{abstract}
Magnetic nanorings have been widely studied due to their potential applications in spintronic and magnonic devices. In this work, by means of analytical calculations and micromagnetic simulations we have analyzed the magnetic energy of nanorings with variable anisotropy along their radius. Four magnetic states, including two new magnetic configurations, here called meron and knot-like states, are considered, looking to the relative lower energy states as a function of anisotropy.  Phase diagrams with this states are presented.
\end{abstract}

\maketitle

\section{Introduction}
Due to  important applications based on the promising concepts of spintronic and magnonic, nanomagnetism has become an  area of intense research in the last decades.  In fact, the production of magnetic nanostructures with different shapes and sizes have been reported in several works \cite{Cowburn-PRD-2000,Streubel-APL-2012,Streubel-Nano-2012,Bacho-NanoTec-2014}. In this context,   magnetic nanorings have become the focus of strong research because of their magnetic behavior. Several works have addressed the  static and dynamic properties of ring?s magnetization  from the theoretical  \cite{Landeros-APL-2007,Bellegia-JMMM,Kravchuck-JMMM,Landeros-JAP,Vagson-JAP-2010,Sebastian-JAP-2014,Riveros-JMMM-2016,Smiljan-JAP-2016,Smiljan-JAP-2017} as well as from the experimental \cite{Klaui-PRL-2001,Klaui-JPC-2003,Zhu-PRL-2006,Vaz-JPC-2007} points of view, . 

Ring-shaped particles are defined by their external and internal radii, $R$ and $r$, respectively, and their thickness, $h$. From magnetic measurements and micromagnetic simulations, three ideal internal magnetic configurations have been identified  in such nanostructures: i) the out-of-plane ferromagnetic state (SD$_\text{z}$); ii) the onion (O); and iii) the vortex (V) configurations \cite{Bellegia-JMMM,Kravchuck-JMMM,Landeros-JAP,Grimsditch-PRB-1998,Abraham-JAP-2001,Hwang-JAP-2000,Cowburn-PRL-1999}. In the SD$_\text{z}$ state, the magnetic moments are parallel to the ring axis, whereas in the other two, the magnetization vector field lays parallel to the ring base. The onion state is accessible from an in-plane saturation of the magnetization and it is characterized by the presence of two opposite walls \cite{Ross-PRB-2003}. In the O and SD$_\text{z}$ states the ring magnetization is nonzero and then the resulting demagnetizing field leads to deviations in the direction of the magnetic moments close to the particle borders, giving rise to edge domains \cite{Lebib-2001}. In the V state, the magnetic moments circulate around the ring axis, and then most of the magnetic flux is confined within the particle. The absence of the highly energetic core region in rings stabilizes the vortex state \cite{Kravchuck-JMMM}, leading to simpler and reproducible switching processes. Consequently, the determination of the conditions for occurring the vortex state in nanorings has been regarded as a key point for the production of new magnetic devices\cite{Chineese,Application-1,Application-2}. 

The resulting magnetization ground-state in nanomagnets is the result of a competition between dipolar, anisotropy and exchange interactions. The phase diagram describing anisotropic nanorings \cite {Landeros-JAP} reveals that magnetization groundstate depends on the relation $R/r$ and the thickness $h$ of the particle. In fact, an increase in the thickness of the nanoring can lead to the formation of a SD$_\text{z}$ state while the increasing in $R$ can favor the formation of O or V states, depending on $r$. Nevertheless, the presence of an uniaxial anisotropy pointing along the ring axis can reduce the thickness for which the SD$_\text{z}$ state becomes the ground-state \cite{Zhang-PRB-2008}.  Therefore, anisotropy could play a very important role in the determination of the magnetization state in a nanoring.

Recently, a large effort has been made to create novel materials with variable magnetic anisotropy, in which the anisotropy could be controlled by electric field \cite{Nicod,Nicoe,Nicof}, temperature \cite{Nicoa,Nicob,Nicoc}, and  ion implantation  \cite{Nicog,Nicoh,Nicoi}. Moreover, it has been shown that reversal magnetization, and domain-wall movement can be controlled by a defined perpendicular anisotropy gradient in Co/Au multilayers using He+ ion-bombardment through a wedged Au stopped layer \cite{Nicoj}. Therefore, a magnetic anisotropy gradient has the potential to lead different domain walls and magnetic states. However, a detailed  study of the effect of an anisotropy gradient
on magnetic nanorings has not been presented yet.

Following these ideas,  the focus of this study  are ring-shaped nanoparticles with variable anisotropy along their radii.  Using analytical calculations and micromagnetic simulations we observe the relative lower energy states of the magnetization, evidencing  two new possible magnetic configurations, here called meron and knot states.

This work is organized as follows: in Section \ref{THModel} we present the theoretical model used in this work. Section \ref{ANResults} brings the obtained analytical results and micromagnetic simulations are presented. Finally, in section \ref{Conclusions} we present the conclusions.

\section{Theoretical model}\label{THModel}
Aiming to determine the magnetization ground state of nanorings with variable anisotropy, we will use a continuous theory in which the magnetization is described as a vector field $\mathbf{M}(\mathbf{r})$ consisting of a smooth function of the position $\mathbf{r}$ inside the magnetic body. For our purposes, the magnetization density will be then written as
\begin{eqnarray}\label{MagProfile}
\mathbf{M}(\mathbf{r})\equiv\mathbf{M}=M_z(\rho,\phi)+M_\phi(\rho,\phi),
\end{eqnarray}
where $\hat{z}$ and $\hat{\phi}$ are unitary vectors in cylindrical coordinates, and $M_z^2+M_\phi^2=M_S^2$, with $M_S$  the saturation magnetization.

We will consider that the dimensions of the nanoring are such that, for vanishing anisotropy,  a vortex configuration is the lower energy state\cite{Landeros-JAP}. For non-vanishing anisotropy, four magnetic configurations are studied. Three of them can be described as $m_{z}=[1-(R_{\bot}-\rho)^2/R_{\bot}^2]^4\cos\kappa\phi$ and $m_\phi=\sqrt{1-m_z^2}$, with $R$ and $r$ the external and internal radii of the cylindrical nanoring, respectively. $R_{\bot}\geq R$ is the limit in which the magnetization points parallel to the $z$-axis in such a way that the larger $R_{\bot}$ the lower the magnetization component  along the $z$-axis direction in the external border of the ring. The value of $R_{\bot}$ will be determined from energy minimization. The three states that can be described by this parametrization are: i) a vortex state (V), obtained in the limit  $R_\bot\gg R$; ii) a meron-like state ($\mathcal{M}$), described as a half-skyrmion configuration \cite{Allende-JMMM-2016}, given by $\kappa=0$; and iii) a knot-like state ($\mathcal{K}$) obtained when  $\kappa\neq 0$ with  the magnetic moments turning around the $z$-axis when the the azimuthal angle is mapped. In this case, $\kappa$ describes how many times the magnetization vector field rotates around the ring axis direction ($z$-axis) in the knot state (See Fig. 1). The fourth magnetization texture  considered is a single domain state (SD$_\text{z}$) in which the magnetic moments  point along the $z$-axis. 

\begin{figure}
\begin{center}
\includegraphics[scale=0.12]{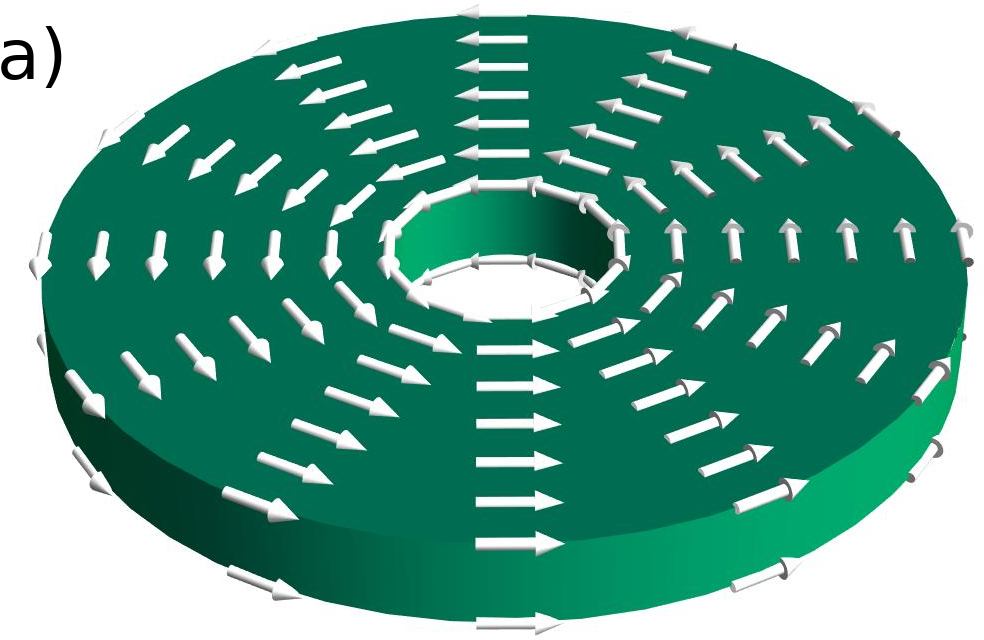}\includegraphics[scale=0.12]{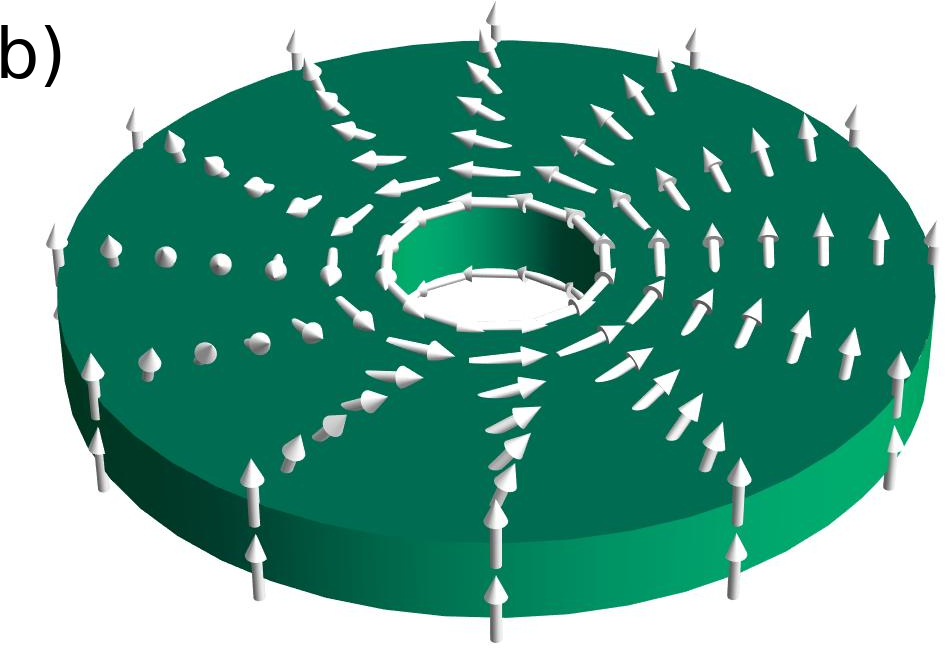}\includegraphics[scale=0.12]{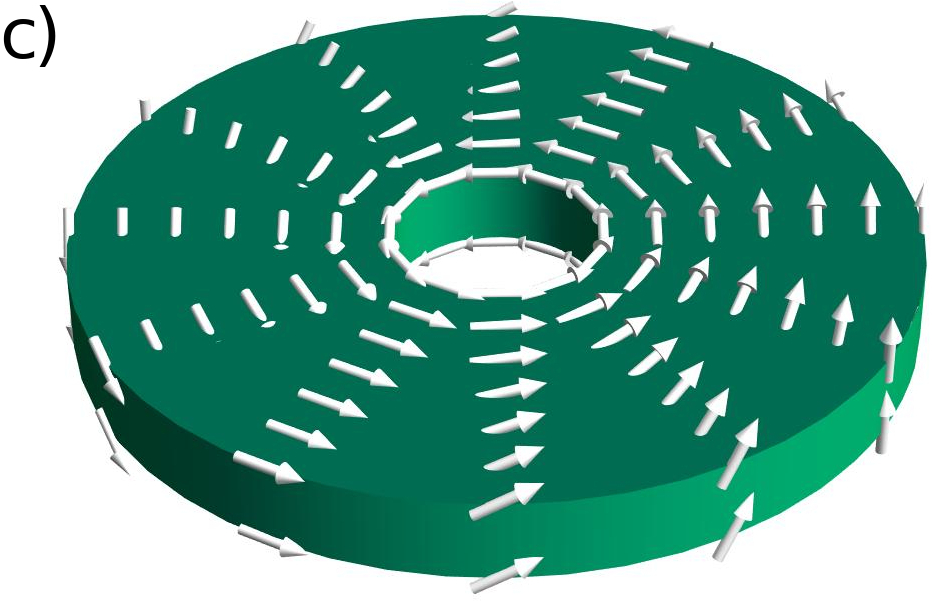}\includegraphics[scale=0.12]{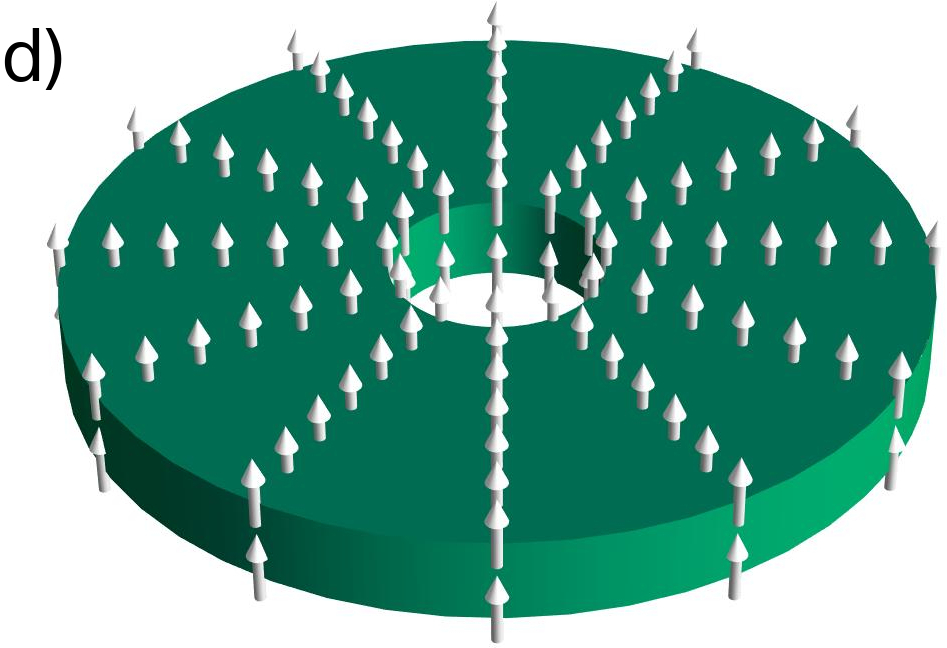}\caption{Magnetic textures a) a vortex, in which all magnetic moments form a close structure. b) a meron-like state, consisting of a vortex state with the magnetic moments gradually pointing upward such that at the external border of the ring $m_z=1$. c) a state with $\kappa=1$, consisting in a smooth transition of the magnetic moments going from -1 at the left of the ring to +1 at its right. d) a SD$_z$ configuration. In all figures, $R_{\bot}=R$}
\end{center}
\end{figure}

In a continuous approach, the  energy ($E_{\text{tot}}$) of a magnetic structure is given by 
\begin{eqnarray}\label{MagneticEn}
E_{\text{tot}}=A\int(\nabla\mathbf{m})^2dV-\int K(\rho)m_z^2\,dV+\frac{\mu_0M_S}{2}\int \mathbf{m}\cdot\nabla U(\mathbf{r})\, dV\,.
\end{eqnarray}
The first, second and third terms in the previous equation correspond to the exchange, dipolar and anisotropy contributions to the magnetic energy, respectively. Here, $A$ is the exchange stiffness, $\mathbf{m}=\mathbf{M}/M_S$, $M_S$ is the saturation magnetization and $K(\rho)$ consists in a  radius-dependent anisotropy constant. 
The exchange energy of the magnetization field described by Eq. (\ref{MagProfile}) is given by
\begin{eqnarray}\label{ExchangeEq}
E_{ex}=Ah\iint\left\{\frac{1}{\rho^2}\left[\frac{1}{1-m_z^2}\left(\frac{\partial m_z}{\partial\phi}\right)^2+1-m_z^2\right]\right.\nonumber\\\left.
+\frac{1}{1-m_z^2}\left(\frac{\partial m_z}{\partial\rho}\right)^2\right\}\rho d\rho d\phi\,.
\end{eqnarray}
It can be notice that if ${m_z}$ does not depend on $\phi$, Eq. (\ref{ExchangeEq}) is reduced to
\begin{eqnarray}\label{ExchangeEqReduced}
E_{ex}=2\pi Ah\int \left[\frac{1}{1-m_z^2}\left(\frac{\partial m_z}{\partial\rho}\right)^2+\frac{1-m_z^2}{\rho^2}\right]\rho d\rho \,
\end{eqnarray} 
according to results presented by Landeros \textit{et al.} \cite{Landeros-PRB-71-2005}. 

\begin{figure}[h!]
\centering
\includegraphics[scale=0.6]{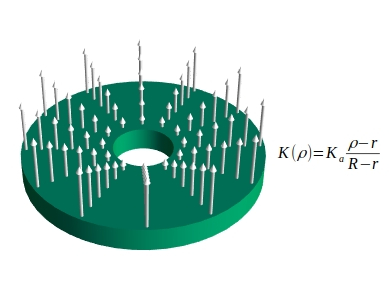}
\caption{Profile of the radius-dependent anisotropy. $K_a$ represents  the anisotropy value at the ring external border}
\label{AnisotropyShape}
\end{figure}

To determine the dipolar energy of the ring, we need to calculate the magnetostatic potential $U(\mathbf{r})$, which can be determined from the surface and volumetric magnetic charges associated with the magnetization. The surface and volumetric magnetic charges are formally defined as $\sigma=\textbf{m}\cdot\mathbf{n}=m_z$ and $\nu=\nabla\cdot\mathbf{m}=m_z(1-m_z^2)^{-1/2}\,\partial_\phi m_z$, respectively. In the absence of currents, the magnetostatic potential can be obtained from solving the Laplace equation, whose formal solution is given by $U(\vec{r})=U_\sigma+U_\nu$, where
\begin{eqnarray}\label{MagnPotSurface}
U_\sigma=\frac{M_S}{4\pi}\left\{\int_{S_1}\frac{\sigma}{|\vec{r}-\vec{r}_1|}dS_1-\int_{S_2}\frac{m_{z}}{|\vec{r}-\vec{r}_2|}dS_2\right\}
\end{eqnarray}
and
\begin{eqnarray}\label{MagnPotVolume}
U_\nu=-\frac{M_S}{4\pi}\int\frac{\nu}{|\vec{r}-\vec{r}'|}dV ,
\end{eqnarray}
are  the magnetostatic potential due to the surface and volumetric magnetostatic charges, respectively. In Eq. (\ref{MagnPotSurface}) $S_1$ and $S_2$ are the surfaces of the top and bottom basis of the cylindrical ring and we have assumed that the ring thickness is very small, in such a way that $m_z$ is constant along the $z$-axis.

In this work we adopt a model in which the magnetic nanoring has an an easy-axis anisotropy gradient pointing along the $\hat{z}$-axis direction and varies linearly with $\rho$. Thus, the anisotropy term varies in the form $K(\rho)=K_a(\rho-r)/(R-r)$, where $K_a$ is the anisotropy at the external border of the ring (see Fig. \ref{AnisotropyShape}).

\section{Results and Discussions}\label{ANResults}
From the described  model, we can calculate the magnetic energy of each  magnetization configuration under consideration. The most simple case is the vortex state, since it has vanishing dipolar and anisotropy energies. Then the magnetic energy of a V state can be evaluated only from the exchange contribution

\begin{eqnarray}
E_{\text{ex}_\text{V}}=2\pi Ah\ln(\frac{R}{r})\,.
\end{eqnarray}

On the other hand, SD$_z$, $\mathcal{M}$ and $\mathcal{K}$ configurations present magnetostatic charges because they have magnetic moments pointing along the $z$-axis  and, consequently, these states present surface magnetic charges. In this context dipolar and anisotropy energies must be taken into account to determine the magnetic energy of such states. Configurations given by $\kappa\geq2$ must demand high exchange energy and then, they will not be considered in this work. Thus, aiming to find the phase space of the SD$_z$, $\mathcal{M}$ and $\mathcal{K}$ states in magnetic nanorings, we will calculate explicitly only the energy associate with $\kappa=0$ and $\kappa=1$. We will start  calculating the exchange energy of such states. Since the magnetization in this case is $\phi$ independent, the exchange energy of the $\mathcal{M}$ configuration is obtained from Eq. (\ref{ExchangeEqReduced}), being evaluated as
\begin{eqnarray}\label{ExEnMS}
E_{\text{ex}_\text{$\mathcal{M}$}}=2\pi Ah\int_r^R\left[64\frac{{m_{z}}^{3/2} }{1-{m_{z}}^2}\frac{(R_\bot-\rho)^2}{R_\bot^4}-\frac{{m_{z}}^2}{\rho^2}\right]\rho d\rho+E_{\text{ex}_\text{V}}\,.
\end{eqnarray}

The $\mathcal{K}$ state given by $\kappa=1$ represents a state in which the magnetic moments turn once along the  $z$-axis direction when going through the nanoring around the azimuthal angle. Due to the azimuthal dependence of the magnetization configuration, the energy of this state must be calculated from Eq. (\ref{ExchangeEq}), giving

\begin{eqnarray}\label{ExEnKSSimp}
E_{\text{ex}_\text{$\mathcal{K}$}}=2\pi Ah\int\left\{\left[64\frac{{m_{z}}^{-1/2} }{\sqrt{1-{m_{z}}^2}}\frac{(R_\bot-\rho)^2}{R_\bot^4}\right]\right.\nonumber\\\left.\times\left(1-\sqrt{1-m_z^2}\right)-\frac{m_z^2}{2\rho^2}-\frac{\sqrt{1-m_z^2}}{\rho^2}\right\}\rho d\rho + 2E_{\text{ex}_{\text{V}}}\,.
\end{eqnarray}

Now, aiming to determine the magnetostatic energy, we can use the Green's function in cylindrical coordinates to describe the inverse of the distance \cite{Morse-book}. After some algebraic manipulation, we can write the magnetostatic potential associated to surface magnetic charges for a generic $\kappa$ as

\begin{eqnarray}\label{SurfacePot}
U_{\sigma}=\frac{M_S\varepsilon_\kappa}{4\pi}\int_{r}^R\rho'd\rho' m_{z}(\rho',\phi')\int_{0}^{\infty}dqJ_\kappa(q\rho)J_\kappa(q\rho')\text{e}^{i\kappa\phi}[\text{e}^{-q(h-z)}-\text{e}^{-qz}]\,,
\end{eqnarray}

\noindent
where $\varepsilon_{\kappa=0}=2\pi$, $\varepsilon_{\kappa\neq0}=\pi$ and $J_\kappa(x)$ is the cylindrical Bessel function of order $\kappa$. The substitution of the previous expression into the dipolar term of Eq. (\ref{MagneticEn}) yields the magnetostatic energy associated to the surface magnetostatic charge, given by

\begin{eqnarray}\label{DipEnSur}
E_{\sigma}=\frac{\mu_0M_s^2\varepsilon_\kappa^2}{4\pi}\int_{0}^{\infty}dq\left[\int_{r}^{R}m_{z}J_\kappa(q\rho)\rho d\rho \right]^2(1-\text{e}^{-hq})\,.\nonumber\\
\end{eqnarray}

\noindent
It can be noticed that if we consider the $\mathcal{M}$ ($\kappa=0$) configuration, the above equation is reduced to the magnetostatic energy of the end-width of a skyrmion \cite{Allende-JMMM-2016}.
%perdon pero que significa the end-width

The calculation of dipolar energy associated to volumetric magnetostatic charges appearing in the $\mathcal{K}$ state can be performed from the evaluation of the magnetostatic potential given in Eq. (\ref{MagnPotVolume}) for a generic value of $\kappa$. In this case, the magnetostatic volumetric charge is evaluated as

\begin{eqnarray}
\nu=\nabla'\cdot \mathbf{m'}=\frac{m_{z}^2\kappa\,\cos\kappa\phi'\sin\kappa\phi'}{\sqrt{1-m_{z}^2\cos^2\kappa\phi'}}\,.
\end{eqnarray}

\noindent
Following the same procedure used to calculate the magnetostatic potential of the surface term, we can expand the inverse of the distance in cylindrical coordinates. Therefore, the volumetric contribution to the magnetostatic potential is given by

%\begin{widetext}
\begin{eqnarray}\label{VolumePotential}
U_\nu=\frac{M_S}{4\pi}\int_{0}^\infty dq\int_{0}^{H}dz'\text{e}^{-q(z_{>}-z_{<})}\int_{r}^R\rho' d\rho'\nonumber\\\times\sum_{m=-\infty}^\infty J_{m}(q\rho')J_{m}(q\rho')\int_{0}^{2\pi}\nu\,\text{e}^{im(\phi-\phi')}d\phi'\,.
\end{eqnarray}
%\end{widetext}

\noindent
In order to simplify our analysis, we  use the property $\cos^2 \kappa\phi'=(1+\cos2\kappa\phi')/2$. Thus, we can calculate the integral in $\phi'$ by using the following series expansion \cite{Morse-book}

\begin{eqnarray}
\frac{1}{\sqrt{1-m_{z}^2\cos^2\kappa\phi'}}=\frac{\sqrt{2}}{m_{z}}\frac{1}{\sqrt{\cosh\eta-\cos 2\kappa\phi'}}\nonumber\\=\frac{2}{\pi m_{z}}\sum_{n=0}^\infty Q_{n-1/2}^0\left(\cosh\eta\right)\cos 2n\kappa\phi'\,,
\end{eqnarray}
where $\cosh\eta\equiv{(2-m_{z}^2)}/{m_{z}^2}$ and $Q^0_{n-1/2}(\cosh\eta)$ is the Legendre function of half-integer order of second kind. In this case, volumetric magnetic charge can be rewritten as
\begin{eqnarray}
\nu=\frac{2\kappa\,m_{z}}{\pi}
\sum_{n=0}^\infty Q_{n-1/2}^0\left(\cosh\eta\right)\cos\kappa\phi'\sin\kappa\phi'\cos 2n\kappa\phi'\,.\nonumber\\
\end{eqnarray}

\noindent 
Substituting the previous expression in Eq. (\ref{VolumePotential}), and assuming that $\kappa$ is an integer and using the properties of integrals of trigonometric functions, we obtain $U_\nu=0$. Therefore, the dipolar energy of $\mathcal{M}$ and $\mathcal{K}$ states can be calculated from Eq. (\ref{DipEnSur}). The magnetostatic energy of SD$_z$ state has been was obtained in Ref. \cite{Landeros-JAP}. 

Finally we determine the anisotropy energy for  the $\mathcal{M}$, $\mathcal{K}$ and SD$_z$ configurations and  arbitrary $\kappa$ values from the third term in Eq. (\ref{MagneticEn}) 
\begin{eqnarray}\label{EAni}
E_{\text{Ani}}=-\varepsilon_\kappa\,h\,K_a\int_r^R\frac{\rho-r}{R-r}\,m_z^2\,\rho\,d\rho\,.
\end{eqnarray}

\noindent
Since in the considered parametrization $\kappa$ is an integer (0 or 1), one can notice that the anisotropy energy of the $\mathcal{K}$ state does not depend on this parameter. Despite the integral in Eq. (\ref{EAni}) has analytical solution, it is cumbersome and will be omitted here.

\begin{figure}[h!]
\centering
\includegraphics[scale=0.1]{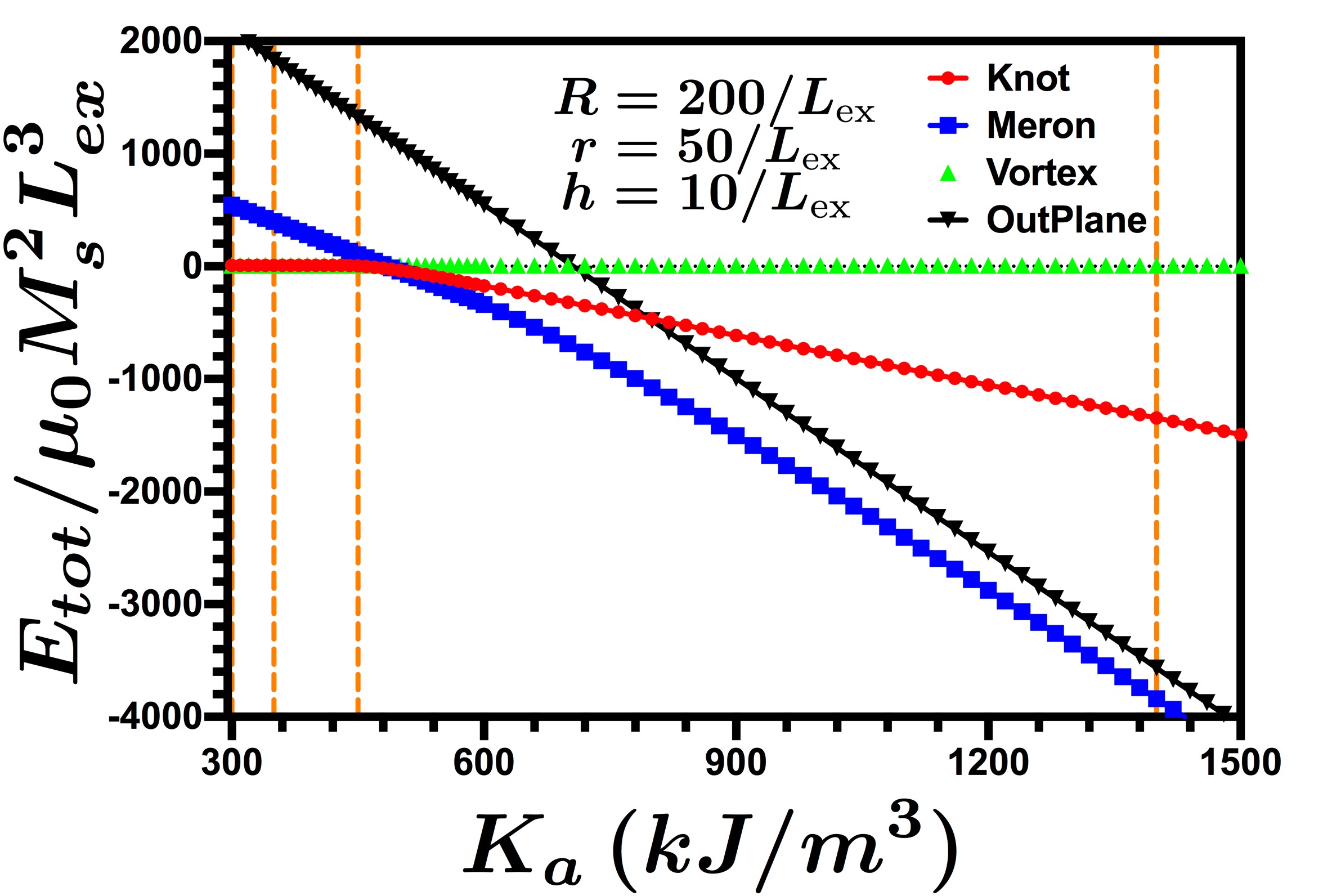}
\caption{Magnetic energy of different magnetic states for Permalloy rings as a function of the anisotropy parameter $K_a$. Results are obtained using $h=20$ nm, $R= 200$ nm and $r=50$ nm. Orange dashed lines represents anisotropy values used to perform micromagnetic simulations.}
\label{Result1}
\end{figure}

The anisotropy contribution for the SD$_z$ state has a simple analytical expression and can be evaluated as
\begin{eqnarray}\label{EAni1}
E_{\text{Ani}}=-\frac{\pi\,h\,K_a}{3}(R-r)(2R+r).
\end{eqnarray}

Aiming to perform a subtle analysis of Eqs. (\ref{ExEnMS}), (\ref{ExEnKSSimp}) and (\ref{DipEnSur}) we have solved the integrals numerically. The nominal magnetic parameters associated to Permalloy are used; that is, $A=1.3 \times 10^{-11}$ J/m, $M_S=8.6 \times 10^5$ A/m, $\mu_0=4 \pi \times 10^{-7}$ J/mA$^2$ and exchange length $\ell_{ex}=5.3$ nm. Such values enables to stabilize all considered magnetization configurations by changing  $K_a$ values. Our results for the energies, considering the different configurations are illustrated in  Fig. \ref{Result1}. These results evidence that, despite the curves for V and $\mathcal{K}$ states are practically superimposed for  $K_a<450$ kJ/m$^3$, there is a small difference in the energies of these magnetic states. In fact, the V state minimizes the energy for $K_a<350$ kJ/m$^3$ but when the anisotropy increases, $\mathcal{K}$ state is the configuration the minimizes the total energy.  Despite the $\mathcal{K}$ state presents lower magnetostatic energy when compared to the $\mathcal{M}$ state, the difference of the anisotropy energy can stabilize the $\mathcal{M}$ configuration for $k_a>700$ kJ/m$^3$. It can be observed that this magnetization pattern is highly stable compared to the SD$_z$ configuration in such a way that SD$_\text{z}$ state does not minimize the total energy in the range of evaluated $K_a$ values. 

To corroborate our analytical results, we have performed micromagnetic simulations using the 3D Object Oriented MicroMagnetic Framework (OOMMF) package \cite{oommf}, which solves the Landau-Lifshitz-Gilbert \cite{LLwork,Gilbertwork} equation using finite element methods.

\begin{figure}
\begin{center}
\includegraphics[scale=0.45]{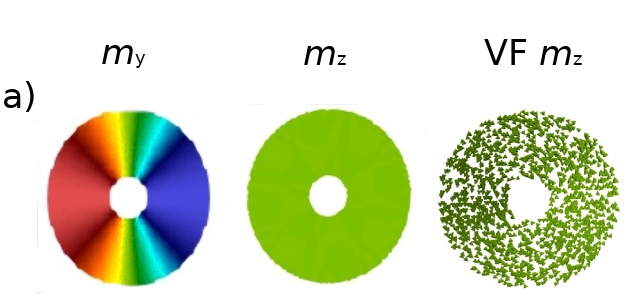}\\\includegraphics[scale=0.45]{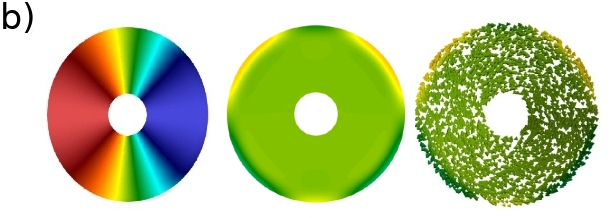}\\\includegraphics[scale=0.45]{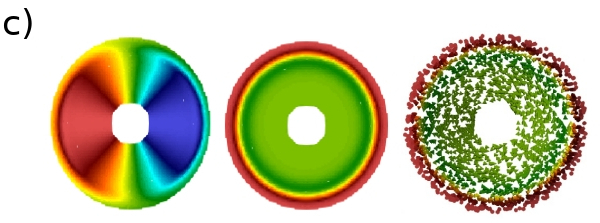}\\\includegraphics[scale=0.45]{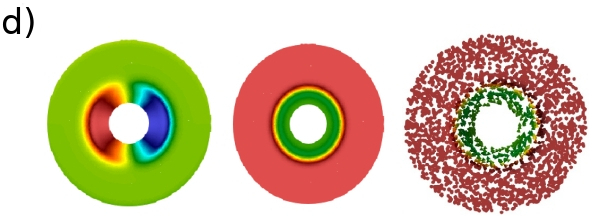}\\\includegraphics[scale=0.45]{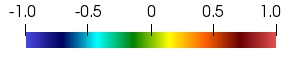}\caption{Final states for different anisotropies from OOMMF simulations. From left to right we show the magnetization components $m_y$, $m_z$ and the vector field (VF). Results obtained for a) $K_a=300$ kJ/m$^3$, b) $K_a=350$ kJ/m$^3$, c) $K_a=450$ kJ/m$^3$ and d) $K_a=1400$ kJ/m$^3$.}\label{Final}
\end{center}
\end{figure}

Simulations were run using a  mesh  size  of  $2\times2\times2$  nm$^3$ and a damping constant of 0.5. Aiming to allow the system to reach equilibrium, we have used a usual torque condition that establishes when  simulations can stop.  That is, the minimum energy state is reached when the torque on the magnetic moments is below 0.001 A/m. Our  simulations were performed  starting from four different initial  configurations:  i) a V configuration; ii) a state in which half of the ring is in a vortex state and the other half with two opposite saturated configurations; iii) a state in which the magnetization in half of the ring is a vortex state and in the other half magnetic moments are pointing along the $z$-axis; and iv) a  single domain pointing along the $xy$-plane. The energy of the final state associated with each initial configuration was calculated, leading us to obtain the lower energy state for each $K_a$ value.

We have then simulated magnetic nanorings with four different anisotropy parameters: 300 kJ/m$^3$, 350 kJ/m$^3$, 450 kJ/m$^3$ and 1400 kJ/m$^3$. The  final states are shown in Fig. \ref{Final}. It can be noticed that the magnetic configuration of the lower energy states agrees with the magnetic configuration of the lower energy state   from analytical calculations. Nevertheless, it can be observed that the final $\mathcal{K}$ state (Fig. \ref{Final}-b) is different from that one described in the analytical model. Indeed, while our analytical model predicts a $\mathcal{K}$ state in which the region with out-of-plane components of the magnetization has the same area that the region with in-plane components, the analysis of Fig. \ref{Final}b shows that the  area occupied by the out-of-plane component of the $\mathcal{K}$ state is larger than the in-plane one. However, the mathematical formalism describing such configuration leads to cumbersome equations and does not present different qualitative changes with the obtained phase diagram. Another important observation coming from the analysis of Fig. \ref{Final}c and \ref{Final}d is that the region of the out-of-plane component to the magnetization for $\mathcal{M}$ state increases with the anisotropy in such a way that for very high values of $K_a$, a SD$_z$ state can be observed. That is, by increasing the anisotropy  the region of the out-of-plane component of the magnetization occupy a larger area of the ring. This finding reveals that the region occupied by the out-of-plane component can be controlled by varying the anisotropy at the border of the ring.

\section{Conclusions}\label{Conclusions}

In this work we have studied the possible magnetic groundstate of a nanoring having an anisotropy gradient along its radius. We have analytically calculated the magnetic energy of four different magnetization configurations and have obtained the energy curves in function of the anisotropy for each considered magnetization configuration. We have also performed micromagnetic simulations and analyzed the  possibility of obtaining two new magnetic configurations in magnetic nanorings, the so called meron and knot-like states. 

The possibility of obtaining the $\mathcal{M}$ and $\mathcal{K}$ states in nanomagnetic systems is very interesting, from the fundamental and the applied points of view. From the fundamental point of view, is always an important issue the appearence of new magnetic textures. From the applied perspectives,  the existence of different magnetic configurations may be associated with new properties that could allow applications in technological developments based on the concept of spintronic and magnonic. Due to emergent new experimental techniques controlling magnetic anisotropy gradient, such new states could be observed.  

We thank Brazilian agency CNPq for finantial support (grant numbers 401132/2016-1 and 301015/2015-5). In Chile we acknoledge Fondecyt under grants 1160198, and the Basal program under grant FB 0807. N. M. Vargas acknowledges the financial support from the Department of Energy's Office of Basic Energy Science under grant DE-FG02-87ER-45332.

\end{document}